\documentclass[prd,aps,twocolumn,reprint,floatfix,nofootinbib,superscriptaddress]{revtex4}
\pdfoutput=1
\usepackage{amssymb,graphicx}
\usepackage{amsmath}
\usepackage{multirow}
\usepackage{epsfig}
\usepackage[usenames]{color}

\newcommand{\beqn}{\begin{eqnarray}}
\newcommand{\eeqn}{\end{eqnarray}}
\newcommand{\beq}{\begin{equation}}
\newcommand{\eeq}{\end{equation}}

\def\mphi{m_{\phi}}

\def\pt{\tilde{p}}
\def\rt{\tilde{\rho}}

\begin{document}

\title{Spontaneous Scalarization with Massive Fields}
\author{Fethi M.\ Ramazano\u{g}lu}
\affiliation{Department of Applied Mathematics and Theoretical Physics, Centre for Mathematical Sciences,\\
University of Cambridge, Wilberforce Road, Cambridge CB3 0WA, UK}
\affiliation{Department of Physics, Ko\c{c} University, \\
Rumelifeneri Yolu, 34450 Sariyer, Istanbul, Turkey}
\author{Frans Pretorius}
\affiliation{Department of Physics, Princeton University, Princeton, New Jersey 08544, USA}

\begin{abstract}
We study the effect of a mass term in the spontaneous scalarization of neutron stars, for
a wide range of scalar field parameters and neutron star equations of state. Even though massless scalars
have been the focus of interest in spontaneous scalarization so far, recent observations of
binary systems rule out most of their interesting parameter space. We point out that adding a mass
term to the scalar field potential is a natural extension to the model that avoids these observational
bounds if the Compton wavelength of the scalar is small compared to the binary separation.
Our model is formally similar to the asymmetron scenario recently introduced in application to cosmology,
though here we are interested in consequences for neutron stars and thus 
consider a mass term that does not modify the geometry on cosmological scales.
We review the allowed values for the mass and scalarization parameters in the theory given current binary 
system observations and black hole spin measurements.
We show that within the allowed ranges, spontaneous scalarization can have nonperturbative, strong effects 
that may lead to observable signatures in binary neutron star or black hole-neutron star mergers, 
or even in isolated neutron stars.
\end{abstract}

\maketitle

\section{Introduction}
Spontaneous scalarization is a phenomenon that occurs in certain
scalar-tensor theories of gravity where the scalar field vacuum can be unstable
to condensation of the field in the presence of certain kinds of matter~\cite{PhysRevLett.70.2220}. 
As we discuss in more detail below, at the linear
level the instability is a long wavelength tachyon instability,
where the minimum wavelength is inversely related to the
density of matter. Non-linear coupling of the scalar field to matter saturates the
growth of the field at a value related to a parameter in the coupling potential.
These facts together allow for the intriguing possibility that, with certain potentials,
amongst all compact matter objects known in the Universe only neutron stars offer an environment
where scalarization can occur. Moreover, the effects can be non-perturbative, allowing order-of-unity
deviation in the structure of neutron stars. Hence this offers a (rare) example of an alternative
theory to general relativity (GR) that is consistent with current weak field observational bounds~\cite{Will:2006LR},
yet could have sizable deviations in the dynamical, strong-field, in particular as pertains
to gravitational wave emission in merger events~\cite{2013PhRvD..87h1506B,2014PhRvD..89h4005S}.

To date, most investigations of spontaneous scalarization have considered massless scalars. However,
recent observations of a pulsar-white dwarf binary~\cite{2013Sci...340..448A} have allowed rather
stringent bounds to be placed on the massless theory, eliminating most of the range of relevance
to future gravitational wave observations. Moreover, as we show below, a massless scalar
would have condensed on cosmological scales, and could be ruled out by cosmological
observations alone\footnote{On the flip side, this can actually be a ``feature'' if one
wants to explain aspects of dark energy or dark matter with scalar tensor gravity, rather than
limit its effects to neutron stars. As far as we are aware two models similar to the 
spontaneous scalarization theory described here are the 
so-called ``symmetron'' \cite{2011PhRvD..84j3521H} and more recently introduced
``asymmetron'' cosmologies~\cite{Chen:2015zmx}; we will discuss similarities with our work
and this latter model further below.}.
A simple way to adjust the theory to evade these observational constraints, yet
preserve the property of giving neutron stars large, non-perturbative corrections to the
predictions of GR, is to give the scalar field a mass $m_\phi$. Such a modification
is also rather ``natural'' compared to the machinations theorists often resort to to conform
their favorite theory to observational contraints (questions about the naturalness of the original 
theory aside). The effect of the mass term is two-fold. First, it suppresses the
tachyon instability for wavelengths longer than the Compton wavelength $\lambda_\phi=2\pi/m_\phi$ of the field
(unless otherwise stated we use units where Newton's constant $G$, the speed of light $c$ and
Planck's constant $\hbar$ are set to 1). Thus a very light mass can prevent the instability
on cosmological scales. Second, it screens the presence of the field outside 
a scalarized neutron star in that the field falls off as $e^{-r/\lambda_\phi}/r$ rather than 
the $1/r$ decay of a massless field. This will effectively shut-off dipole radiation
in a white dwarf-neutron star system if the orbital radius is significantly larger 
than $\lambda_\phi$~\cite{2012PhRvD..85f4041A}; 
it is the lack of inferred dipole radiation that allows the observations presented in~\cite{2013Sci...340..448A} 
to so tightly constrain the massless theory.

Motivated by the above considerations, in this work we present an initial study of massive scalar field 
spontaneous scalarization in neutron stars. Our main goal is to investigate the static solutions representing isolated
non-spinning neutron stars within this theory, for various parameters of the theory and neutron star equations of state
(EOS). This is a first step toward exploring the mergers of binary neutron star and black hole-neutron
star systems, which we are currently pursuing and will present the results of elsewhere.  
Much of what has been discussed above about scalarization is well known. Independent of 
our work, a mass term has recently been discussed~\cite{Chen:2015zmx} in what the authors dub the asymmetron 
scenario. While the primary motivation for the asymmetron is cosmological,
we are interested in a modified theory which differs from GR only on small scales relevant
to the late stages of compact object coalescence involving neutron stars.
We will discuss the differences between our model 
and the asymmetron in the results section. 

To our knowledge, no detailed work on
neutron star structure or binary mergers have been performed for massive field 
scalarization in the fully nonlinear regime. We aim to start this discussion by understanding 
the properties of isolated scalarized neutron stars.
As such, after introducing the theory in Sec.~\ref{sec_boe} we give
back-of-the-envelope calculations
illustrating the properties discussed above. This suggests masses in the range 
$10^{-15} {\rm eV}\lesssim m_\phi\lesssim10^{-9}{\rm eV}$
are consistent both with present observational constraints, yet can produce non-perturbative
deviations in the structure of neutron stars. A mid-section 
$10^{-13} {\rm eV}\lesssim m_\phi\lesssim10^{-11}{\rm eV}$ of this range can further be eliminated
if putative measurements of the spin of solar mass black holes are accurate~\cite{Narayan:2013gca},
and these are old black holes. The reason is if a rapidly spinning
black hole forms with Schwarschild radius close to the Compton wavelength of a massive scalar field,
superradiant amplification of the scalar field will occur, causing the black hole to loose most of its
angular momentum on timescales sufficiently short to make observation of the initially highly spinning
black hole unlikely~\cite{2013PhRvL.111k1101C}.

In Sec.~\ref{sec_res} we solve the analog of the
Tolman-Oppenheimer-Volkov (TOV) equations in this theory to find the static neutron star 
solutions. We search over the parameter space of the scalar tensor theory to 
see where scalarization in neutron stars occur, and ascertain its effect on the star's structure.
We show that for $\lambda_\phi$ much larger than the radius of the neutron star, the mass term (unsurprisingly)
has little effect compared to the massless theory. 
For $\lambda_\phi$ much smaller than the radius of the neutron star, the mass term prevents scalarization,
and the usual neutron stars of GR result. Scalarization can lead to large observable effects in certain parts
of the parameter space even for isolated stars. We will discuss the implications of such effects and how future gravitational wave 
observations of binary mergers could constrain this theory.

\section{Equations of Motion and the Origin of Spontaneous Scalarization.}\label{sec_boe}
The Lagrangian of the scalar-tensor theory that leads to spontaneous scalarization is given by~\cite{PhysRevLett.70.2220}
\begin{align}
 \frac{1}{16\pi} &\int d^4x \sqrt{g} \left[R - 2g^{\mu \nu} \partial_{\mu} \phi  \partial_{\nu} \phi
 - 2 m_{\phi}^2 \phi^2 \right] \nonumber \\
 &+ S_m \left[\psi_m, A^2(\phi) g_{\mu \nu} \right]
\end{align}
where $g_{\mu\nu}$ is the metric in the Einstein frame, $\phi$ is the scalar field and $m_{\phi}$ is 
the parameter coupling to the mass potential. $S_m$ is any ordinary matter contribution 
to the Lagrangian with the matter degrees of
freedom represented by $\psi_m$. $\psi_m$ couple to a conformally scaled version of the metric, 
$\tilde{g}_{\mu\nu} = A^2(\phi) g_{\mu\nu}$, rather than the minimal coupling in GR.
The scaled metric defines the so-called Jordan frame, and is the physical metric observers use to measure
proper length-scales. We use a tilde to denote any tensor defined in this frame.
The equations of motion are
\begin{align}\label{EOM}
 R_{\mu\nu} &= 8\pi \left( T_{\mu\nu} -\frac{1}{2} g_{\mu\nu} T \right) + 2 \partial_{\mu} \phi \partial_{\nu} \phi
 + \mphi^2 \phi^2 g_{\mu\nu} \nonumber \\
 \Box_g \phi &= -4 \pi \alpha(\phi) T + \mphi^2 \phi
\end{align}
where $\alpha = \partial \left(\ln A \right) / \partial \phi$ and $\Box_g$ is the wave 
operator with respect to the Einstein frame metric. We use $A(\phi)=e^{\beta \phi^2/2}$
throughout this study, with $\beta$ a constant parameter, but other choices are also possible\footnote{Aside from the 
coefficient of the parabolic $\phi$ term, $\beta$, another important property of the 
potential $A$ is its asymptotic value $A_\infty = A(\phi \to \infty)$. This parameter 
determines the deviation from GR for extremely strong scalar fields. Our choice of $A_\infty = 0$ 
gives the maximal possible deviation, whereas values closer to $1$ set an {\em a-priori} upper 
limit on the observable differences from GR. The asymmetron scenario considers such 
nonzero $A_\infty$ values.}.
We will only consider negative values as in~\cite{PhysRevLett.70.2220}

Note that $\phi=0$ is a solution in this theory, and is equivalent to GR. Spontaneous
scalarization occurs when this solution is unstable, i.e. an arbitrarily small perturbation of 
$\phi$ grows and the system ends up in a stable configuration with nonzero $\phi$.

First, let us give a sketch of the physical mechanism behind spontaneous 
scalarization (see~\cite{PhysRevLett.70.2220,2014PhRvD..89h4005S,Chen:2015zmx})
for alternative approaches). To relate the dynamics of the field
to physical properties of matter, we rewrite the scalar field equation of motion
as follows
\begin{equation} \label{scalar_eom}
  \Box_g \phi = -4 \pi \beta e^{2\beta \phi^2} \phi \tilde{T} + m^2_{\phi} \phi
\end{equation}
where  $\tilde{T}$ is the trace of the physical stress-energy tensor. Beginning with
a small perturbation about the GR solution $\phi=0$, we can expand to linear order in $\phi$:
\begin{equation} \label{sf_lin_eom}
\Box_g \phi \approx  \left( - 4 \pi \beta \tilde{T} + m^2_\phi \right)\phi
\end{equation}
For matter that can be modeled as a perfect fluid, $\tilde{T}=-\tilde{\rho}+3\tilde{P}$,
where $\tilde{\rho}$ and $\tilde{P}$ is the (physical) rest-frame density and pressure
of the fluid respectively. For non-relativistic matter, $\tilde{\rho}\gg\tilde{P}$,
$\tilde{T}\approx-\tilde{\rho}$, and so for $\beta<0$ the first term on the right in (\ref{sf_lin_eom})
is effectively a negative mass-squared term. At the linear level the theory thus suffers a tachyon
instability where $\lambda_{eff}<\lambda_\phi$, with
\begin{equation}\label{star_bl}
\lambda_{eff}\equiv\sqrt{\frac{\pi}{|\beta|\tilde{\rho}}}.
\end{equation}
Consequently all Fourier modes with 
wavelength $\lambda > \lambda_{eff}/\sqrt{1-(\lambda_{eff}/\lambda_\phi)^2}$ that ``fit'' within 
the region where $\lambda_{eff}<\lambda_\phi$ 
will initially experience exponential growth. 
This of course by itself would be disastrous for the theory, though from (\ref{scalar_eom})
one can see that the non-linear term $e^{2\beta \phi^2}$ will eventually become important and 
suppress the growth, saturating $\phi$ at a value, order of magnitude, of $1/\sqrt{|\beta|}$.

For a star, approximating $\tilde{\rho}\sim M/R^3$, where $M$ is the mass of the star and $R$ its radius,
\begin{equation}\label{lambda_star}
\lambda_{eff,star}\sim~\frac{R}{\sqrt{C|\beta|}}, 
\end{equation}
with $C\equiv 2M/R$ being its compactness. 
To be susceptible to scalarization, $\lambda_{eff,star}<\lambda_\phi$,
and the shortest wavelength unstable mode must fit inside the star, or roughly 
$R>\lambda_{eff,star}$.
Thus, for a given $\beta$, only stars that are sufficiently compact
\begin{equation}\label{C_b}
C \gtrsim 1/|\beta|
\end{equation}
can scalarize. For a typical $1.4M_\odot$ neutron star $C$ is approximately between $1/5$ and $1/3$ (depending
on the equation of state), a white dwarf has $C\sim10^{-3}$, and a main sequence solar mass star has $C\sim 10^{-6}$.
Note however, that for very massive neutron stars, and again depending on the EOS, the
core can become relativistic in that $\tilde{T}\gtrsim 0$, which will suppress scalarization.

On cosmological scales, during the radiation dominated epoch $\tilde{T}\approx 0$, however in the
matter dominated era $\tilde{T}\approx-\tilde{\rho}_m$, where $\tilde{\rho}_m$ is the average, 
redshift $z$-dependent energy density in matter. Thus, unless the scalar field has a sufficiently large mass
term, the entire Universe would scalarize (see also the discussion in~\cite{Sampson:2014qqa}). 
To estimate how large a mass term is required
to prevent this, let us assume that the non-relativistic component of matter became
relevant at matter-radiation equilibrium $z_{eq}\approx 10^3$. Then, $\tilde{\rho}\equiv \tilde{\rho}_{m,eq} \approx\tilde{\rho}_{m0} z_{eq}^3$,
where ${\rho}_{m0}\sim 3\times 10^{-27}{\rm kg}/{\rm m}^3$ is the present day baryonic matter density. 
Relative to the matter density at $z_{eq}$, to prevent scalarization would thus require
\begin{equation}
\lambda_\phi \lesssim 10^5 {\rm pc} \sqrt{\frac{\tilde{\rho}_{m,eq}}{|\beta| \tilde{\rho}_m}},
\end{equation}
or
\begin{equation}\label{m_cos_est}
m_\phi \gtrsim 10^{-27} {\rm eV} \sqrt{\frac{|\beta| \tilde{\rho}_m}{\tilde{\rho}_{m,eq}}}.
\end{equation}
Note that a similar order of magnitude estimate could have been obtained using
the earlier analysis for stars, since considered a uniform density sphere the compactness
of the Universe reaches unity for a radius of order the Hubble length.
If a tachyonic instability were excited in the Universe, a simple estimate for its
growth rate can be obtained solving (\ref{sf_lin_eom}) on a flat background and assuming
$\rho_m$ is constant : $\phi\propto e^{a t}$, with $a=\sqrt{4\pi |\beta| \tilde{\rho}_m}$. 
Scaled to $z_{eq}$, this gives an e-folding time of 
\begin{equation}
t_s\sim 10^6{\rm yr} \sqrt{\frac{\tilde{\rho}_{m,eq}}{|\beta| \tilde{\rho}_m}}.
\end{equation}
Thus unless $|\beta|$ is extremely small scalarization happens very rapidly on a cosmological timescale.
Though note that regardless of the magnitude of $\beta$, once the instability saturates
the effect is an order unity scaling between the Einstein and Jordan frame 
metrics ($\tilde{g}_{\mu\nu}=e^{4\beta\phi^2} g_{\mu\nu}$, $\phi\propto 1/\sqrt{|\beta|}$).

Returning to stellar sources,
far from the star, the spherically
symmetric static solution for the case with a mass term is
\begin{equation} \label{scalar_asymptotic_massive}
\phi(r \to \infty) \sim a \frac{e^{-2\pi r/\lambda_{\phi}}}{r} \ .
\end{equation}
For the massless case, this changes radically to
\begin{equation} \label{scalar_asymptotic_massless}
\phi(r \to \infty) \sim \phi_{\infty} + \frac{a}{r} \ ,
\end{equation}
where $\phi_{\infty}$ and $a$ are constants. Thus it is apparent
that the mass term effectively screens the potential on scales larger
that $\lambda_\phi$, and also removes the ambiguity in the vacuum state
of the field (i.e. $\phi_{\infty}$=0).

\subsection{Theoretical and observational bounds on the parameters of the theory}
Our main theoretical motivation for considering spontaneous scalarization is to explore
the consequences of an alternative theory of gravity to GR that (i)
is consistent with GR in all regimes where it has been tested by
experiment or observation, (ii) predicts large deviations from GR in the dynamics and 
consequently gravitational wave emission during strong-field merger events, and (iii)
has a classical, mathematically well-posed initial value problem. There are several reasons why these
restrictions are important.
First is to understand the issue of theoretical bias in the gravitational wave detection
effort~\cite{Yunes:2009ke}. This can arise due to the heavy reliance on theoretical templates for gravitational wave observation:
if GR does not exactly describe the dynamical strong-field regime
relevant to the late stages of merger, unless templates are used that explicitly measure
this, the result could likely be detections erroneously attributed to pure GR with ``wrong''
parameters for the binary. Various methods, such as the parameterized post Einsteinian (ppE) 
approach~\cite{Yunes:2009ke},
have been proposed to try to measure the
consistency of a signal with GR, though without explicit examples beyond perturbation
theory for how the waveforms can differ, it is unclear how effective these approaches may be.
Case in point is the ``dynamical scalarization'' effect noted in binary neutron star
mergers within the massless theory in~\cite{2013PhRvD..87h1506B}, where at close separations prior to merger a scalarized 
neutron star is able to induce scalarization in its initially un-scalarized companion; this affects
the waveform in a manner not well captured by the original ppE parameterization.
Second, it is unclear whether using a theory that violates (i) is useful, even if only to use as a strawman to measure
the effects of deviations from GR; i.e. consistency in the weak-field and with the leading order
radiative dynamics of GR may in general severely constrain possible deviations in the strong-field.
Lastly, if a theory violates (iii), aside from the obvious doubts that would place on its viability,
it will not be solvable using standard numerical methods. It is rather remarkable that (to our knowledge)
scalar tensor theories with spontaneous scalarization are the only class of alternatives that
have been demonstrated to satisfy (i), (ii) and (iii).

These considerations thus guide the following choice of parameters within the theory that we consider
viable. First, the observational constraints inferred from the pulsar-white dwarf binary 
PSR J0348+0432~\cite{2013Sci...340..448A}
have come close to ruling out essentially the entire range of parameters in the massless theory that lead to neutron stars
being scalarized. The massless theory also necessarily affects the Universe on cosmological
scales. We thus require a mass term (see~\cite{Chen:2015zmx} for an alternative cosmological view). To avoid bounds 
from PSR J0348+0432 requires that the mass be sufficiently large such that $\lambda_\phi \ll r_p$,
with $r_p\sim10^{10}{\rm m}$ is the periapse of the orbit (the actual orbital parameters
of the binary are very accurately measured, though here we just give the order of magnitude 
results).
This translates to a lower limit for the mass
\begin{equation}
m_\phi \gg 10^{-16}{\rm eV},
\end{equation}
which also easily suppresses cosmological effects\footnote{The asymmetron scenario proposes 
a lower bound of $10^{-11} \mathrm{eV}$~\cite{Chen:2015zmx}. This arises from a 
completely different consideration motivated by cosmology.}.
From (\ref{C_b}), to allow a star as compact as a neutron star
to scalarize, but not a white dwarf, bounds $\beta$ to the range
\begin{equation}
3 \lesssim -\beta \lesssim 10^3.
\end{equation}
To not have the mass term prevent scalarization
in a neutron star requires $\lambda_{eff,star} < \lambda_\phi$, which depends
on the structure of the star (see (\ref{lambda_star})). For a strict upper limit
consider a neutron star where $C|\beta|$ is maximal ($C\sim1/3$ and $|\beta|=10^3$); 
with $R\approx 10{\rm km}$ this gives
\begin{equation}
m_\phi \lesssim 10^{-9}{\rm eV}.
\end{equation}
For $|\beta|$ approaching its lower limit, this upper limit on $m_\phi$ increases by about two orders of magnitude.

A further restriction on the scalar mass can be leveled if claimed measurements
of high spins for several candidate stellar mass black holes are correct~\cite{Narayan:2013gca}.
The reason is highly spinning black holes are superradiantly unstable
in the presence of a massive scalar field with Compton wavelength on order the
size of the black hole~\cite{2013PhRvL.111k1101C}. The effect of the instability
is to spin down the black hole; thus observation of a highly spinning, old black hole
rules out existence of a related range of scalar field masses. Taking the present
observations of black hole spins as doing so rules out the mass window from
roughly  $10^{-11}\ \textrm{eV}$ to $10^{-13}\ \textrm{eV}$.

\section{results}
With the above guidance on restrictions to the scalar-tensor theory parameters, we investigate
the spontaneous scalarization scenario for various equations of state and various values of
$\beta$ and $\mphi$. 
\subsection{TOV Framework.}
We seek the static solutions for perfect fluid neutron stars in the spontaneous
scalarization theory with a mass term; the massless-case study was originally carried out in~\cite{PhysRevLett.70.2220}.
We use the following Einstein-frame ansatz for the metric:
\begin{equation}
g_{\mu\nu} dx^{\mu} dx^{\nu} = -e^{\nu(r)} dt^2 + \frac{dr^2}{1-2\mu(r)/r} + r^2 d\Omega^2.
\end{equation}
In terms of physical quantities, the perfect fluid stress energy tensor is
\begin{equation}
\tilde{T}^{\mu\nu}=(\rt+\pt)\tilde{u}^{\mu}\tilde{u}^{\nu}+\pt \tilde{g}^{\mu\nu},
\end{equation}
where the energy density $\rt$, pressure $\pt$, and components of the fluid 4-velocity $\tilde{u}^{\alpha}$
only depend on the radial coordinate $r$
(and due to the symmetries $\tilde{u}^{\alpha}=e^{-\nu/2}(\partial/\partial t)^{\alpha}$).
The equations of motion~(\ref{EOM}) reduce to the following set of ordinary differential equations (ODEs):
\begin{align}
 \mu' &= 4\pi r^2 A^4(\phi) \rt + \frac{1}{2}r(r-2\mu) \psi^2 + \frac{1}{2} r^2 \mphi^2 \phi^2 \nonumber \\
 \nu' &= r\psi^2 + \frac{1}{r(r-2\mu)}\left[r^3[8\pi A^4(\phi)\pt-\mphi^2\phi^2] +2\mu\right] \nonumber \\
 \phi'&= \psi \nonumber \\
 \psi' &(r-2\mu) = 4\pi r A^4(\phi)\left[\alpha(\phi)(\rt-3\pt)+r\psi(\rt-\pt)\right] \nonumber \\
 &\ \ \ \ \ \ \ \ \ \ \ \ \  +\mphi^2 (r^2\phi^2\psi +r\phi)-2\psi(1-\mu/r) \nonumber \\
 \pt' &= -(\rt+\pt)\left( \nu'/2+\alpha(\phi) \psi \right)
\end{align}
where $'$ denotes a derivative with respect to $r$. 
This system of equations is closed by supplying
an equation of state of the form $\rt=\rt(\pt)$. Then, to solve the equations requires specifying
initial conditions at $r=0$, and integrating outwards. At the surface of the star
the pressure goes to zero, and beyond this point we set $\rt$ and $\pt$ to zero,
integrating only the scalar field and metric equations further outward.
Regularity at the origin requires $\mu(0)=\nu(0)=\psi(0)=0$. In general one can freely specify
$\pt(0)=\pt_0$ and $\phi(0)=\phi_0$, for which the
asymptoptic solution for the scalar field of an isolated star
takes the form $r\phi(r)=Ae^{-2\pi r/\lambda_\phi}+Be^{2\pi r/\lambda_\phi}$. Only solutions
with $B=0$ are physically relevant, and hence for a given $\pt_0$, if a scalarized solution
exists it will correspond to a particular (non-zero) value of $\phi_0$.
We numerically find these solutions 
using the \textit{shooting method}. This begins with a guess for values of
$\phi_0$ that bracket $B=0$, then using a bisection
search to reduce the range to $|B|<\epsilon$, for some predetermined small tolerance $\epsilon$.
Of course, for any finite $B$ the solution will eventually blow up, though if sufficiently
small it will match the $B=0$ solution to a desired accuracy for $r<r_1$, with $r_1$ a
chosen outer boundary location. We use a $4th$-order Runge-Kutta method to integrate the ODEs.
A further issue for the $B=0$ solutions is a sub-class of them turn out to be dynamically
unstable to perturbations. As we will describe later, to investigate this
we evolved a set of neutron stars using the numerical code described 
in~\cite{Pretorius:2004jg,code_paper} (imposing axisymmetry).

\section{Results}\label{sec_res}
We investigate the existence and the strength of spontaneous scalarization for different 
values of $\beta$ and $m_{\phi}$, as well as different EOS. We use
the piecewise-polytropic parameterization for the EOS introduced in~\cite{2009PhRvD..79l4033R}, 
designed to approximate the zero temperature limit of many of
the current nuclear-physics-inspired EOS, and bracket much of the theoretically plausible
range. These equations of state named 2B, B, HB, H, 2H correspond to successively increasing stiffness for the 
neutron star matter, 2B being the softest, and 2H the stiffest. We measure the strength of the scalar field 
by its value at the center of the neutron star, where it is maximal, and also discuss 
its dependence on the radius for various scalar field parameters.

\begin{figure*}
\begin{center}
\includegraphics[width=5in,draft=false]{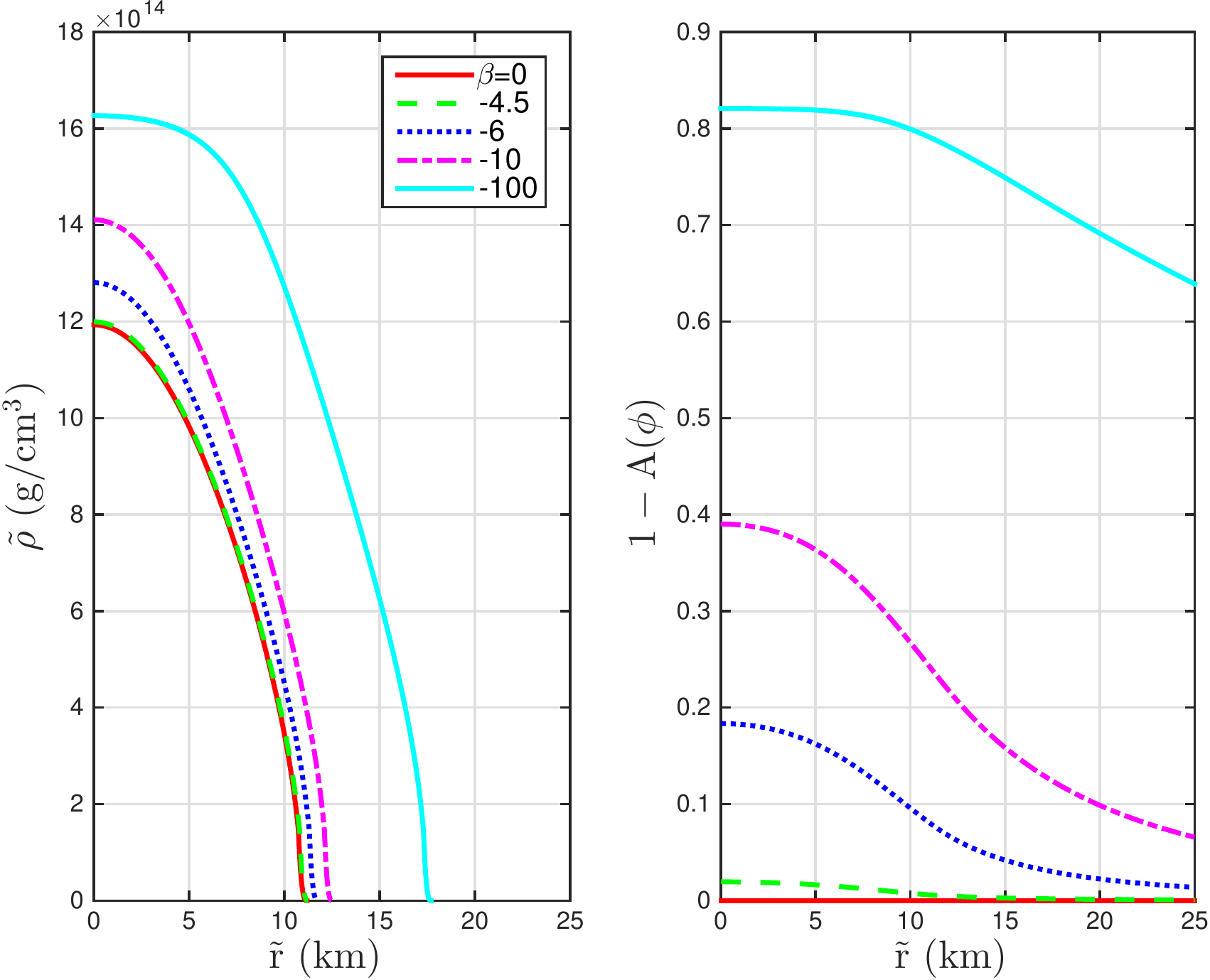}
\end{center}
\caption{
Effect of varying $\beta$ on the radial profiles of the matter density (left) and 
conformal factor $1-A(\phi)$ (right) for neutron stars.
The physical radius $\tilde{r} = A(\phi(r)) r$ is the radial 
coordinate associated with the Jordan frame metric. All cases have a fixed ADM mass of 
$1.70 M_{\odot}$, $m_\phi=1.6\times 10^{-13} \mathrm{eV}$ and HB EOS.  The baryon mass of the 
neutron star is $2.06, 2.07, 2.10, 2.32, 5.37 M_{\odot}$ for $\beta=0, -4.5, -6, -10, -100$ 
respectively. Even for moderate values of $\beta$, the structure of the star is altered significantly. 
For large values such as $\beta=-100$, observations of isolated stars might already be able to
test spontaneous scalarization.
\label{star_profile_beta}
}
\end{figure*}

\begin{figure*}
\begin{center}
\includegraphics[width=5in,draft=false]{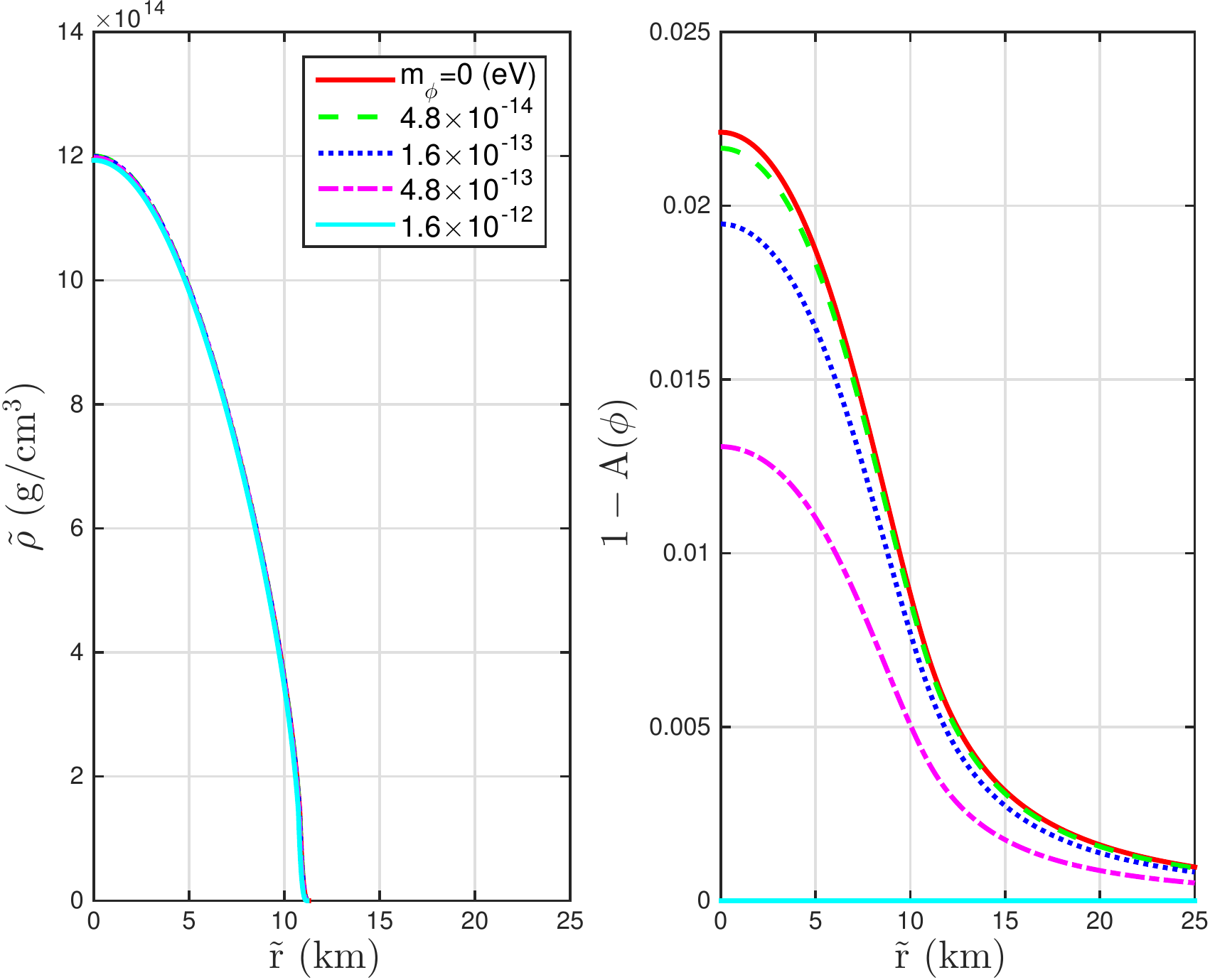}
\end{center}
\caption{
Effect of varying $m_\phi$ on the radial profiles of the matter density (left) and 
conformal factor $1-A(\phi)$ (right) for neutron stars.
The physical radius $\tilde{r} = A(\phi(r)) r$ is the radial coordinate 
associated with the Jordan frame metric. All cases have a fixed ADM mass of 
$1.70 M_{\odot}$, $\beta=-4.5$ and HB EOS. Increasing $m_\phi$ always inhibits the 
scalarization of the star and eventually kills it. Below a certain value 
($m_\phi \sim 10^{-13} \mathrm{eV}$ for this case) the field profile changes 
only marginally, asymptoting to the case of a massless scalar. This characteristic 
dependence on $m_\phi$ holds qualitatively for higher $\beta$ as well, 
but the cutoff value of $m_\phi$ that allows scalarization is typically higher for 
more negative $\beta$. The baryon mass of the neutron star is within a 
percent of $2.06 M_{\odot}$ for all cases. 
\label{star_profile_mass}
}
\end{figure*}

Radial profiles of the density and scalar field of a spherically symmetric neutron star with representative
values of $\beta$ and $m_\phi$, and for the HB equation of state, are shown in 
Figs.~\ref{star_profile_beta} and \ref{star_profile_mass}. 
For the cases with stronger scalar fields, the structure of the star changes noticeably, 
especially near the center where the deviation from general relativity is greatest. 
Also, the campactness of the star can be significantly different for higher values of $|\beta| \gtrsim 10$.

\begin{figure*}
\begin{center}
\includegraphics[width=6in,draft=false]{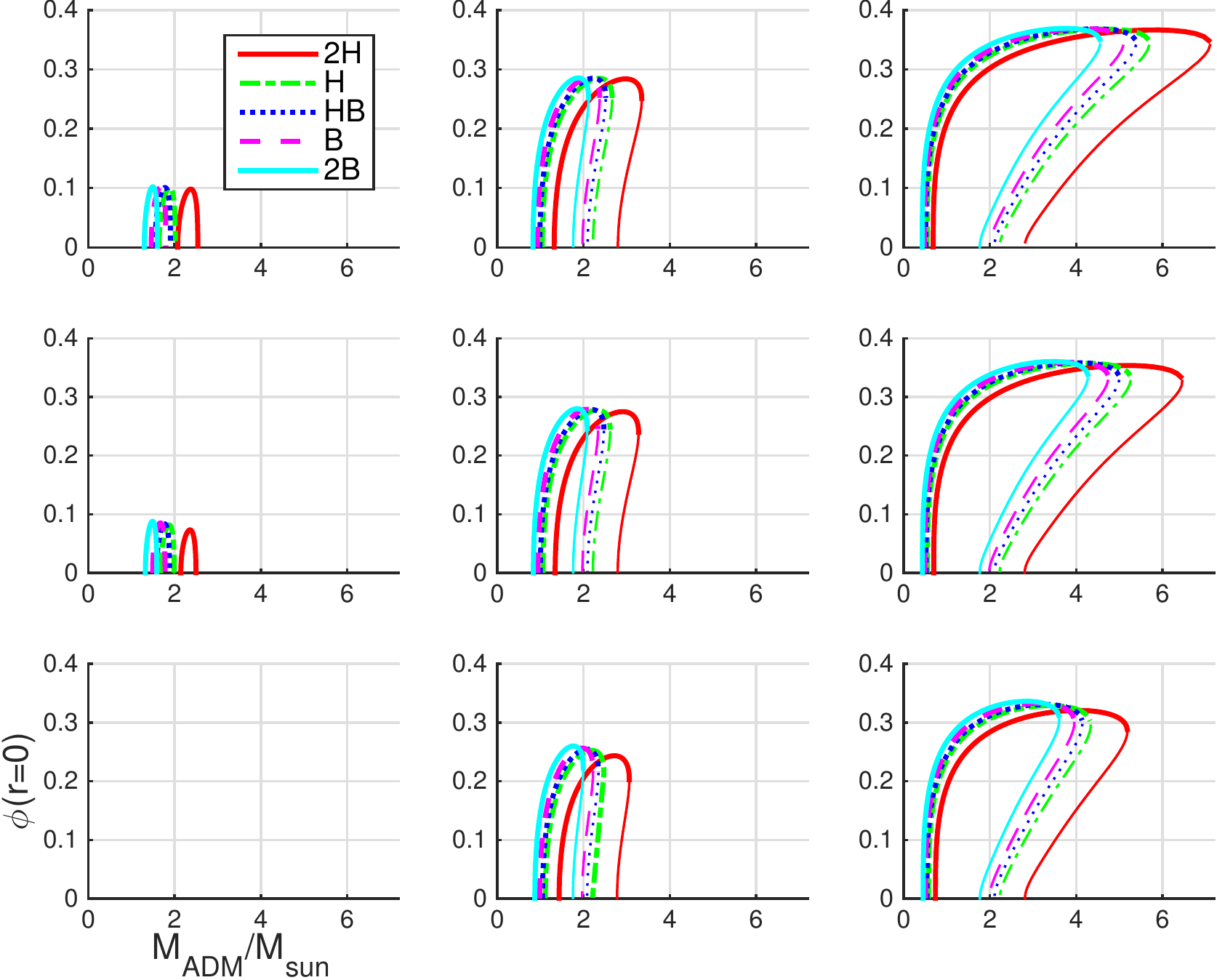}
\end{center}
\caption{
Maximum value of the scalar field for different representative values of $\beta$
and scalar field mass $m_{\phi}$. Each subplot shows the maximum scalar field value 
with respect to the ADM mass of the neutron star for various EOSs described 
in~\cite{2009PhRvD..79l4033R}, $\textrm{2H}$ is the stiffest and $\textrm{2B}$ is the 
softest. Upper row: $m_{\phi}= 1.6 \times 10^{-13}  \textrm{eV}$, middle 
row: $m_{\phi}= 4.8 \times 10^{-13}$, lower row: $m_{\phi}= 1.6 \times 10^{-12} \textrm{eV}$, 
left column: $\beta =-4.5$, middle column: $\beta=-6$, right column: $\beta=-10$. 
When two scalar field values are possible for a given ADM mass, the solution corresponding
to the lower scalar field is unstable (thinner lines on the right end of the curves).
Spontaneous scalarization becomes weaker with increasing $m_{\phi}$ and decreasing 
$\beta$, and eventually disappears (lower left). 
Note that the allowed maximum ADM mass of scalarized neutron stars can be quite large 
compared to the GR maximum mass even for moderately negative values of $\beta$.
\label{parameter_space}
}
\end{figure*}

The strength of spontaneous scalarization for several different points in $\beta-m_\phi$ parameter space 
and for various EOS is illustrated in Fig.~\ref{parameter_space}. For a given point in parameter space
(i.e one of the subplots in Fig.~\ref{parameter_space}) and a given EOS, there is a finite ADM mass range for 
the star that allows spontaneous scalarization. Even though the EOS can affect the neutron star mass range 
where spontaneous scalarization can occur, it does not have a strong effect on its qualitative behavior, i.e. 
the existence of the effect and the maximum strength of the scalar field has a comparatively weak dependence on 
the EOS compared to varying $\beta$ and $m_\phi$.

One interesting feature of the dependence of the scalar field strength on the ADM mass is that in certain
cases two different scalar field profiles are possible for a given ADM mass. Anticipating that
only one of these solutions is dynamically stable, we evolved these stars using the code
described in~\cite{Pretorius:2004jg,code_paper}, and found that the star 
with the lower $\phi(r=0)$ value is unstable in all cases (the thin parts of the curves in Fig.~\ref{parameter_space}). These 
solutions quickly evolve to either a stable configuration with similar ADM mass, or collapse to a black hole. 
This is analogous to behavior seen in boson star solutions within pure GR, e.g~\cite{Hawley:2000dt}.
All unstable configurations we investigated have more than one extrema as a function of radius, whereas the stable stars have monotonically decreasing scalar field profiles.

\subsection{Discussion}

Figs.~\ref{star_profile_beta}-\ref{parameter_space} clearly demonstrate that increasing $m_{\phi}$ weakens, and eventually prevents spontaneous scalarization for all $\beta$, which is to be expected from (\ref{sf_lin_eom}). Theoretical studies of the massless theory have already shown that the maximal value of the spontaneously scalarized field (or total scalar charge) drops by a few orders of magnitude around $\beta=-4.5$, and then disappears completely around $\beta=-4$ for a wide variety of EOS~\cite{2013PhRvD..87h1506B,2014PhRvD..89h4005S}. This is consistent with the estimate (\ref{C_b}), and still qualitatively true for massive scalars (see Fig.~\ref{parameter_space}). Although this is not surprising for low $m_{\phi}$, where the TOV-like equations are small perturbations of the massless scalar case near the star, it holds true even near the upper scalar mass limits imposed by the radius of the neutron star. In short, a scalar mass term does not significantly alter the least negative value
of $\beta$ below which spontaneous scalarization occurs, as compared to the massless theory. For even more negative $\beta$ values, neutron stars can again support significant scalar fields for a wide range of $m_{\phi}$ values, with an upper bound that depends on the radius of the star. 

On the other hand, as discussed in the introduction, even though the range of $\beta$ that allows strong spontaneous scalarization is similar for the massive and massless cases, a significant difference appears once observations are used to restrict the parameter space. Namely, more negative $\beta$ values ($\lesssim -4.5$) that lead to strong scalarization for a massless scalar but which are ruled out by the white dwarf-pulsar binary~\cite{2013Sci...340..448A} are still viable for a massive scalar field, 
due to its suppression of dipole radiation for large binary separations.
However, significantly more negative values of $\beta$ start to induce radical changes to the structure
of a star that might be constrained by observations of single neutron stars. For example, at fixed ADM mass, scalarization leads to a more than $50\%$ increase in the stellar radius around $\beta \sim -100$ (see Fig.~\ref{star_profile_beta}). 
Such large radii could likely be ruled out by existing observations of thermonuclear bursts from neutron
stars (see e.g.~\cite{2015arXiv150505155O}). Though to properly connect these observations to inferred mass/radii require models
of the bursts and subsequent light propagation within the geometry of the star, and how this is affected by
scalarization. This would be an interesting line of inquiry for a future study.

We also note that the effect of scalarization on the star's structure depends strongly on the EOS, but always {\em increases} the maximum allowed ADM mass. This is especially evident for the softest 2B EOS, which cannot support a neutron star with the highest observed mass under GR\cite{2013Sci...340..448A, 2010Natur.467.1081D}, but can do so for $\beta \lesssim -5$.

Another avenue to test spontaneous scalarization is via gravitational wave observations of compact object mergers involving neutron stars. We expect the dynamical scalarization effect~\cite{PhysRevD.89.044024}, the phenomena where the strong scalar field of one neutron star can induce scalarization in a companion which in isolation would not carry a significant field, to become less pronounced as the mass of the field increases. This is due to the exponential decay
of the field outside the star. 
As scalarization also allows significantly higher ADM mass values for $\beta \lesssim -10$ (see Fig.~\ref{parameter_space}), if that is the case a larger fraction of binary neutron star mergers 
will leave a massive neutron star remnant versus a black hole. These starkly contrasting outcomes
will produce very different gravitational wave signals after coalescence. We plan to pursue
some of these directions of study in follow-up work.

Lastly, we give brief comments contrasting our results to the asymmetron model~\cite{Chen:2015zmx}, focusing on consequences for scalarized stars. The asymmetron imposes $m_\phi \gtrsim 10^{-11}\mathrm{eV}$ and $\beta \ll -1$. However, a direct comparison in terms of these 
two parameters is not straight-forward, as the asymmetron model imposes a different limiting
value on the conformal scaling function $A$, namely that $A_\infty =A(\phi \to \infty)$ is a
positive, order-of-unity value not close to $0$. $A_\infty =1$ corresponds to GR, while $A_\infty =0$ is the strongest possible
deviation from GR, as in our model. Thus some of the radical changes mentioned above to neutron star structure
for very large, negative $\beta$ values can be ameliorated in the asymmetron model by varying the asymptotic behavior of $A$.
However, in general (without more careful ``engineering'' of the conformal scaling function $A$), parameters in our model
designed to give large but viable deviations to compact object physics are not relevant to physics on cosmological scales, and
vice versa. 

\section{\em Conclusions}
A significant feature of the original spontaneous scalarization scenario was that it was immune to existing
weak field and binary pulsar constraints, allowing for the intriguing possibility that large deviations
to the structure of neutron stars were possible compared to the predictions within pure general relativity.
The more recently discovered white dwarf-pulsar binary system has now almost ruled out this massless version
of spontaneous scalarization. In this work we pointed out that the addition of a mass term
to the scalar field potential can restore this feature of spontaneous scalarization without being 
in conflict with these observations. 

Our preliminary calculations show that roughly a five order of magnitude range
for the scalar field mass $m_{\phi}$ is viable. We computed the static solutions for isolated neutron stars
for representative values of  $m_{\phi}$ within this range, showing that spontaneous
scalarization exists and can be strong (depending on the coupling parameter $\beta$). 
Our primary goal for exploring this theory is to have a well-posed vehicle to
explore deviations to general relativity in the dynamical strong-field, of relevance
to the late stages of binary compact object coalescence. As such this study is a first
step toward studies of merger simulations of scalarized binary neutron star and black
hole neutron star systems in the massive theory, which we plan to pursue in future work.

\acknowledgments
This research was supported by NSF grants PHY-1065710, PHY-1305682, NASA grant NNX11AI49G,
STFC GR Roller Grant No. ST/L000636/1 (FMR) and the Simons 
Foundation (FP). Computational resources were provided by the {\it Orbital} cluster at Princeton 
University and the {\it COSMOS} Shared Memory system
at DAMTP, University of Cambridge operated on behalf of the STFC
DiRAC HPC Facility. The latter equipment is funded by BIS National
E-infrastructure capital grant ST/J005673/1 and STFC grants
ST/H008586/1, ST/K00333X/1, ST/J001341/1.

We thank Vitor Cardoso, Luis Lehner and Ulrich Sperhake for useful discussions.

\bibliographystyle{h-physrev}
\bibliography{st}

\end{document}